\documentclass[prd,aps,floats,onecolumn,twoside,preprintnumbers,
superscriptaddress,floatfix,nofootinbib,showpacs]{revtex4}
\usepackage{graphicx,colordvi,pstricks,rotating,amsmath}
\usepackage{amssymb,epstopdf,bbm}
\usepackage{slashed}
\usepackage[T1]{fontenc} 
\usepackage{epsfig}

\begin{document}

\title{Revisiting NCQED and scattering amplitudes}

\author{Josip Trampeti\'{c}}
\affiliation{Ru\dj er Bo\v{s}kovi\'{c} Institute, Division of Experimental Physical, Bijeni\v{c}ka 54, 10000 Zagreb, Croatia}
\email{josip.trampetic@irb.hr}
\affiliation{Max-Planck-Institut f\"ur Physik, (Werner-Heisenberg-Institut), F\"ohringer Ring 6, D-80805 M\"unchen, Germany}
\email{trampeti@mppmu.mpg.de}
\author{Jiangyang You}
\affiliation{Ru\dj er Bo\v{s}kovi\'{c} Institute, Division of Physical Chemistry, Bijeni\v{c}ka 54, 10000 Zagreb, Croatia}
\email{jiangyang.you@irb.hr}
 
\newcommand{\tr}{\hbox{tr}}
\def\BOX{\mathord{\vbox{\hrule\hbox{\vrule\hskip 3pt\vbox{\vskip
3pt\vskip 3pt}\hskip 3pt\vrule}\hrule}\hskip 1pt}}
 
\date{\today}  

\begin{abstract}


Research progresses on the noncommutative gauge theories on the Moyal space are discussed in this minireview.  We first present a brief overview on the development of gauge theories on Moyal space, with an emphasis on the role of Seiberg-Witten maps. Two important relations induced by reversible Seiberg-Witten maps, namely the formal equivalence of on-shell DeWitt background field effective action in general and explicit identical relation between tree level scattering amplituded in noncommutative quantum electrodynamics (NCQED), are described in some details. We then proceed to the properties of the tree level two-by-two scattering amplitudes in NCQED, including a forward scattering singularity in NCQED Compton scattering. After covering some phenomenological perspectives of NCYM based models, outlooks into the future are given at the end.

\end{abstract} 

\maketitle

\section{Introduction}

While held perfectly in contemporary experimental physics, possible causes of Lorentz symmetry violation at higher energy scales have been repeated proposed for various reasons. In search for theory of Quantum Gravity (QG) within a braneworld scenario of string theories \cite{ArkaniHamed:1998rs}, it was pointed out that distant astrophysical objects with rapid time variations could provide the most sensitive opportunities to probe very high energy scales, i.e., almost the near-Planck scale physics ($E=1.2 \times 10^{19}$ GeV) \cite{AmelinoCamelia:1997gz}. Thus there is a number of generic Quantum Gravity frameworks involving a  spontaneous breaking of Lorentz symmetry, when a tensor field acquires a vacuum expectation value (vev). Unlike the case of scalars, these tensor vevs do carry spacetime indices, causing interactions represented by the Standard Model (SM) fields coupled to these vevs to depend on the direction or velocity of the said fields. Stated differently, these background vevs bring up deformations that break the Lorentz symmetry.

Deformation mentioned above could be introduced via spacetime noncommutativity and by the quantization of the electromagnetic field in such spacetime. That was an original Heisenberg idea, motivated to cure unavoidable UV divergences of the quantum field theory (QFT),  suggested to Peierls and shared with Pauli and Oppenheimer, who's student Snyder realised it and published a bit later \cite{Snyder:1946qz,Snyder:1947nq}. Simple field theory formulated on such deformed space-time is implemented by replacing the usual pointwise product of the ordinary fields $\phi(x)$ and $\psi(x)$ by the star($\star$)-product in any action, follows:
$\phi(x)\psi(x) \longrightarrow (\phi\star \psi)(x)=\phi(x)\psi(x) + {\cal O}(\theta,\partial \phi,\partial \psi).$
The specific Moyal-Weyl $\star$-product is relevant for the case of $\theta^{\mu\nu}$ tensor and it is defined as\footnote{The $\star$-product has also an alternative integral formulation, making its non-local character more transparent.}:
\begin{equation}
(\phi\star \psi)(x)
=e^{\frac{i}{2}\theta^{\mu\nu}{{\partial}^\eta_\mu}\,
\,{{\partial}^\xi_\nu}} \phi(x+\eta)\psi(y+\xi)\big\vert_{\eta,\xi\to0}
\equiv\phi(x) e^{\frac{i}{2}\overleftarrow{{\partial}_\mu}\,
\theta^{\mu\nu}\,\overrightarrow{{\partial}_\nu}} \psi(x).
\label{f*g}
\end{equation} 
Even though that idea did not progress for many years, the NC gauge field theories were defined in analogy to Yang-Mills (YM) theory with matrix multiplication replaced by the Moyal-Weyl $\star$-products. The resulting noncommutative Yang-Mills (NCYM) action is invariant under NC gauge transformations. Such models appear quite naturally in certain limits of string theory in the presence of a background $B^{\mu \nu}$ field. Specifically, a low-energy limit is identified where the entire boson-string dynamics in a Neveu-Schwartz condensate is described by Seiberg and Witten  \cite{Seiberg:1999vs} as a minimally coupled supersymmetric gauge theory on NC space such that the mathematical framework \cite{Kontsevich:1997vb} of NC geometry/field theory \cite{Madore:2000en,Jurco:2000fb,Jurco:2001rq,Jurco:2001kp,Jurco:2001my,Jackiw:2001jb,Madore,Gomis:2000zz,Aharony:2000gz} does apply. In such a scenario, Dirac-Born-Infeld action \cite{Dirac:1928hu,Dirac:1931kp,Born:1934gh} is realized on the NC space as a special limit of open strings in a background $B^{\mu \nu}$ field, in which closed string (i.e. gravitational) modes are decoupled, leaving only open string interactions. Since in string theory $B^{\mu \nu}$ field is a rather mild background, the constant antisymmetric deformation tensor $\theta^{\mu \nu}$ governing noncommutativity of spacetime is not specified, and therefore the scale $\Lambda_{\rm NC}$ could lie anywhere between the weak and the Planck scale \cite{Szabo:2009tn}. In a model of the NC spacetime we consider coordinates $x^\mu$ as the hermitian operators $\hat x^\mu$, satisfying 
$[\hat x^\mu ,\hat x^\nu]=[x^\mu \stackrel{\star}{,} x^\nu]=i\theta^{\mu\nu}$,
\cite{Jackiw:2001jb}, $\vert\theta^{\mu\nu}\vert\sim\Lambda^{-2}_{\rm NC}$, ${\rm dim [\theta^{\mu\nu}]}=[mass^{-2}]=[length^2]$. Also it is straightforward to realize that $\theta^{\mu\nu}$-matrix is related to the Maxwell field strength tensor $F^{\mu\nu}$ since NCQED arises from string theory in the presence of background electromagnetic fields. It is important to set a bound on the $\Lambda_{\rm NC}$ from experiments \cite{Szabo:2009tn}. 

NCYM theories on Moyal space can be quantized perturbatively following the standard BRST protocols~\cite{Martin:1999aq}. Subsequent quantum field theory calculations have been active research topics for over two decades now. Some recent progresses are going to be summarized in this minireview. In the next section we briefly describe the historical developments of NCYM modeling, with an emphasis on the role of Seiberg-Witten (SW) maps. We then summarize the recent studies on scattering amplitudes of NCYM models defined via SW maps in Section 3. Recent phenomenological progresses of NCYM models are discussed in Section 4 before we give some discussion and make conclusion.

\section{Moyal NCYM as quantum field theory: recent years}

Thanking to the associativity and integral cyclicity of the Moyal $\star$-product~\eqref{f*g}, it is relatively straightforward to formulate a gauge filed theory valued in $n\times n$ complex matrix space, which coincides with the universal enveloping algebra of the Lie-algebra U(N), and produces NCYM action as
\begin{equation}
S=\int\frac{-1}{4g^2}\;{\rm Tr}\;F_{\mu\nu}\star F^{\mu\nu},\;
F^{\mu\nu}=\partial^\mu A^\nu - \partial^\nu A^\mu 
-i[A^\mu\stackrel{\star}{,}A^\nu].
\label{NCYM}
\end{equation}
This action allows an elegant perturbative BRST quantization~\cite{Martin:1999aq}, which set the basis of studying Moyal-deformed NC gauge theories as perturbative quantum field theories.

\subsection{NC gauge field theories and Seiberg-Witten maps}

One essential issue of the action~\eqref{NCYM} from the D-brane effective action perspective is that the underlying $\rm D_p$-brane permutation symmetry should induce a commutative U(N) gauge symmetry instead of the $\star$-product gauge symmetry, which requires a way to connect these two symmetries. A few other issues include that an ordinary gauge field $a_\mu(x)$ transforms like a vector under a change of coordinates, $a'_\mu(x')=\frac{\partial x^\nu}{\partial x'^\mu}a_\nu(x)$, while for the noncommutative gauge field $A_\mu(x)$ this holds only for rigid, affine coordinate changes \cite{Jackiw:2001jb}. The possible choice of charges for NC gauge field $A_\mu$ to couple to a matter field are also restricted to ($0,\pm1$) times a fixed unit of charge.

The issues above were, eventually, all solved by celebrated SW maps \cite{Seiberg:1999vs} between ordinary and the noncommutative fields/quantities. SW map promote not only the noncommutative fields and composite nonlocal operators of the commutative fields, but also the noncommutative gauge transformations as the composite operators of the commutative gauge fields and gauge transformations \cite{Liu:2000mja,Okawa:2001mv,Brace:2001rd,Brandt:2003fx,Martin:2002nr,Barnich:2002pb,Horvat:2011qn,Martin:2012aw,Trampetic:2015zma}. Moyal deformed NCYM models were defined for arbitrary gauge group representations via SW maps after it was realized that the NC fields must be valued in the universal enveloping algebra of the commutative gauge Lie-algebra while the commutative gauge fields do not have to be valued in the commutative U(N) Lie-algebra~\cite{Barnich:2002pb,Martin:2012aw}.

This approach to the NC theories avoiding both the gauge group and the U(1) charge issues has been established for quite some time \cite{Madore:2000en,Jurco:2001rq}. It was shown mathematically rigorously that any U(1) gauge theory on an arbitrary Poisson manifold can be deformation-quantized to the NCQFT via such enveloping algebra approach~\cite{Jurco:2001kp} and later extended to the non-Abelian cases \cite{arXiv0711.2965B,arXiv0909.4259B}. 

Building semi-realistic NC deformed particle physics models are made easier with the help of SW maps. Serious efforts on formulating NCQFT models, with phenomenological influence, strongly boosted by the SW map based enveloping algebra approach enables a direct deformation of comprehensive phenomenological models like the noncommutative Standard Model (NCSM) or Grand Unified Theories (NCGUT), respectively. So it appear to study ordinary gauge theories with additional couplings induced by the SW map/deformation. To include a reasonably relevant part of all SW map induced couplings, one usually calls for an expansion of the action in powers of 
$\theta^{\mu\nu}$ \cite{Calmet:2001na,Behr:2002wx,Aschieri:2002mc,Martin:2013gma}, inspired by the original work of Seiberg and Witten \cite{Seiberg:1999vs} leading to truncated NCQFT with respect to deformation parameter $\theta^{\mu\nu}$. One-loop quantum properties of such truncated theories, as well as studies of some new physical phenomena, like breaking of Landau-Yang theorem, was performed. It was also observed that allowing a deformation-freedom via varying ratios between individual gauge invariant terms could improve the renormalisability of such quantum field theory at one-loop level~\cite{Buric:2006wm,Latas:2007eu,Buric:2007ix,Martin:2009sg,Martin:2009vg,Buric:2010wd,Hewett:2000zp,Godfrey:2001yy,Schupp:2002up,Minkowski:2003jg,Ohl:2004tn,Alboteanu:2006hh,Buric:2007qx,Garg:2011aa}.

Several methods exist for deriving solutions of the SW maps~\cite{Seiberg:1999vs,Mehen:2000vs,Liu:2000mja,Okawa:2001mv,Jurco:2001my,Brace:2001rd,Barnich:2002pb,Martin:2012aw}. These solutions are highly nonlinear and not unique. Their nonuniqueness can be understood as a local gauge freedom in this context~\cite{Jurco:2001my,Jurco:2001kp}, which may help to understand the background (in)dependence of the string theory. Finally  note that SW maps also relate Morita-equivalent star products on Poisson manifolds.\footnote{We are particularly grateful to Peter Schupp for comments on the connection between SW maps and the Morita equivalence among star products and on the Kontsevich formality approach \cite{Kontsevich:1997vb} (see the next subsection).}

\subsection{Formal equivalence of NCYM before and after reversible SW map}

Unlike slightly more sophisticate $\kappa$-Minkowski or Snyder spaces~\cite{Grosse:2005iz,Meljanac:2011cs,Meljanac:2017grw,Meljanac:2017jyk}, perturbative quantum field theories on Moyal space is relatively easy to formulate~\cite{Filk:1996dm}. NCYM theories on Moyal space can be quantized perturbatively following the standard BRST method~\cite{Martin:1999aq}. A number of novel results arise from the subsequent loop calculations in these theories. The most significant one is probably the existence of infrared (IR) divergences in one loop two point function~\cite{Minwalla:1999px,Matusis:2000jf,VanRaamsdonk:2000rr,Hayakawa:1999yt,VanRaamsdonk:2001jd,Horvat:2011bs,Ferrari:2003vs,Ferrari:2004ex}, which are often named as ultraviolet/infrared (UV/IR) mixing because the quadratic IR divergence in the scalar $\phi^4$ theory on Moyal space came from part of the UV divergent integral of the commutative theory~\cite{Minwalla:1999px}.\footnote{Note that UV/IR mixing is also connected to the holography in the model independent way \cite{Horvat:2010km} and nicely implemented  into the idea of scalar fields weak gravity conjecture, where it manifests itself as a form of hierarchical UV/IR mixing \cite{Lust:2017wrl}.} It is worthy to record that the leading IR divergences are still quadratic in NCYM, despite the fact that UV divergences are regarded as logarithmic. The leading UV and IR divergences in the supersymmetric NCYM (NCSYM) theories on Moyal space do match each other.

Note also that we have successfully constructed the 3D Aharony–Bergman–Jafferis–Maldacena  (ABJM) theory on the Moyal noncommutative supersymmetric space and show that all UV and IR divergences in 2- and 3-point functions of the theory disappear \cite{Martin:2017nhg}. 

The equivalent loop calculations for SW mapped gauge theories took many years to develop due to the complicatedness of SW map solutions. The first nontrivial advancement in this line of research was the realization that the SW-mapped actions must be formulated $\theta$-exactly~\cite{,Zeiner:2007,Schupp:2008fs}, motivated by the earlier works on $\theta$-exact/closed solutions of SW maps using Konstevich formality map~\cite{Jurco:2000fb,Jurco:2001rq,Jurco:2001my,Jurco:2001kp}, open Wilson lines~\cite{Liu:2000mja,Mehen:2000vs}, and D-brane based methods~\cite{Okawa:2001mv}. Over the next few years, loop integral basis for new integrals induced by SW map were derived for one loop two point function calculations in U(1) NCSYM~\cite{Horvat:2011bs,Horvat:2011qg,Horvat:2013rga,Horvat:2015aca,Martin:2016zon}. A Seiberg-Witten differential equation based order by order solution to the $\theta$-exact SW map for arbitrary gauge group representations was also developed~\cite{Martin:2012aw}. 
A formal equivalence between NCYM before and after reversible SW map was then proven at the level of on-shell DeWitt background field effective action by realizing that the perturbative expansion of the functional Jacobian induced by SW map is a sum of vanishing loop integrals~\cite{Martin:2016hji,Martin:2016saw}. Explicit matches of one loop two point functions in U(1) NCSYM were found to occur only when the gauge fixing and subtraction of equation of motion are set to be exactly the same in the loop calculations of both theories, with and without SW map, respectively~\cite{Martin:2016hji,Martin:2016saw}. Otherwise off-shell artefacts can take place~\cite{Martin:2016zon}. Thus, the DeWitt on-sheel background field effective action should be considered as the appropriate frameworrk to compare (other) quantized NCYM theories in general.

The quadratic IR divergences are still active within the one loop DeWitt on-sheel background field effective action. Therefore we consider it as an object of physical relevance. Nowadays the existence of NC IR divergences in QFT on Moyal space can be considered as firmly established. There are also evidences suggesting their existence in other NC spaces like Snyder space~	\cite{Meljanac:2017jyk}. Full understanding of its impact on these theories, especially NCYM, is still an open problem in this field.

\section{Revisit NCQED scattering amplitudes}

Because of the cumbersome four field interactions induced by SW map, tree-level two-by-two scattering amplitudes of NCQED after SW map were only evaluated directly recently~\cite{Horvat:2020ycy,Latas:2020nji,Trampetic:2021awu}. The results lead to a simple identical relation between NCQED before and after reversible SW map~\cite{Trampetic:2021awu}. A few properties of these scattering amplitudes, which are considerably overlooked in early researches are also noticed, and going to be summarised below.

\subsection{The $\theta$-exact NCQED first model based on reversible SW maps}

We start with the following minimal/first $\theta$-exact NCQED model acion in terms of one noncommutative gauge field and one left charged fermion field, $A^\mu$ and $\Psi$, respectively~\cite{Madore:2000en}:
\begin{equation}
\begin{split}
&^1S[A^\mu(a_{\rho})]
=\int\frac{-1}{4e^2}\;F_{\mu\nu}[A^\mu(ea_{\rho})]\star F^{\mu\nu}[A^\mu(ea_{\rho})]
+\bar\Psi(\psi, e a_\rho)(i\slashed{D}-m)\Psi(\psi, e a_\rho),
\\
&D^\mu\Psi=\partial^\mu\Psi - i A^\mu\star\Psi,
\end{split}
\label{NCminAction}
\end{equation} 
with self-evident notations for the Moyal-Weyl $\star$-commutator. 
The SW map for explicit calculations was derived as the $\theta$-exact expansion with respect to the formal power of commutative field $a_\mu$ from the SW differential equations~\cite{Martin:2012aw}. Brutal force calculations of all tree-level two-by-two scattering processes were performed with respect to dynamical fields $a_\mu$ and $\psi$~\cite{Horvat:2020ycy,Latas:2020nji}. The results showed that scattering amplitudes of all tree-level two-by-two processes coincides exactly with the same model without SW map~\cite{Latas:2020nji}. By inspecting the cancellation of SW map contributions in the two-by-two scattering amplitudes, we realized that there is a systematic cancellation between the SW map extensions of interaction vertices in specific Feynman (sub-)diagrams of the scattering amplitudes. This observation led to the conclusion that all tree-level scattering amplitudes of the first model with SW map coincide with the same model without SW map~\cite{Trampetic:2021awu}.

\subsection{Two-by-two scattering amplitudes}

Scattering amplitudes of the NCYM without SW map were studied in the early years of Moyal NCQFT~\cite{Hewett:2000zp,Godfrey:2001yy}. Some remarkable results were obtained for NCYM without SW map thanking to the vast progresses in researches on scattering amplitudes of commutative gauge theories~\cite{Raju:2009yx,Huang:2010fc}. Our most recent results suggest that scattering amplitudes of NCQED with and without SW maps are connected, therefore it is still relevant to better understand the scattering amplitudes of NCQED without SW map. Here we summarize some of our experiences on this subject.

\subsubsection{Four photon scattering}

An elegant result of the U(N) NCYM scattering amplitudes is that there are the color/star-product ordered amplitudes of NCYM identical to their counterparts in commutative SU(N) gauge theories~\cite{Raju:2009yx,Huang:2010fc}. The NC factors only enter the scatterinng ampltiudes at the resummation stage. For this reason, the NCQED scattering amplitudes are naturally connected with the SU(N) gauge theories like QCD. An example of such connection could be realized by the fully helicity summed scattering amplitude square of the four photon scattering ($\gamma\gamma\to\gamma\gamma$) process~ \cite{Hewett:2000zp,Latas:2020nji}
\begin{equation}
\begin{split}
&\sum\limits_{helicity}\big\vert{\cal M}(\gamma(k_1)\gamma(k_2)\to\gamma(k_3)\gamma(k_4))\big\vert^2
=2\Big(\left\vert M_{\rm NC}^{++++}\right\vert^2+\left\vert M_{\rm NC}^{++--}\right\vert^2
\\&
+\left\vert M_{\rm NC}^{+-+-}\right\vert^2\Big)
=256 e^4\cdot(-)
\bigg[\sin^2\frac{k_1\theta k_2}{2}
\sin^2\frac{k_3\theta k_4}{2}\left(\frac{t}{u}+\frac{u}{t}+\frac{tu}{s^2}\right)
\\&
+\sin^2\frac{k_1\theta k_3}{2}\sin^2\frac{k_2\theta k_4}{2}\left(\frac{s}{t}+\frac{t}{s}+\frac{st}{u^2}\right)
\\&
+\sin^2\frac{k_1\theta k_4}{2}\sin^2\frac{k_2\theta k_3}{2}\left(\frac{s}{u}+\frac{u}{s}+\frac{su}{t^2}\right)\bigg].
\end{split}
\label{Helicity25}
\end{equation}
The conversion to QCD amplitude square can be achieved by replacing the NC factors with identical color summation factors 
\begin{equation}
\sin^2\frac{k_i\theta k_j}{2}\sin^2\frac{k_k\theta k_l}{2}\;
\longrightarrow\;\sum\limits_{\{\alpha_{i},\alpha_{j},\alpha_{k},\alpha_{l}\}}\Big(\sum\limits_{\mu}f^{\alpha_i\alpha_j\mu}f^{\alpha_k\alpha_l\mu}\Big)^2,
\label{Helicity27}
\end{equation}
where $i\not=j\not=k\not=l=1,2,3,4$.
One can then sum over the Mandelstam variable fractions and obtain
\begin{eqnarray}
(-)\left(\frac{t}{u}+\frac{u}{t}+\frac{tu}{s^2}+\frac{s}{t}+\frac{t}{s}\right.
\left.+\frac{st}{u^2}+\frac{s}{u}+\frac{u}{s}+\frac{su}{t^2}\right)
=3-\frac{su}{t^2}-\frac{st}{u^2}-\frac{tu}{s^2},
\label{Helicity28}
\end{eqnarray}
which coincides exactly with the QCD amplitude square of $gg\to gg$ process (Peskin \& Schroeder Eq.(17.78) \cite{Peskin:1995ev}).

On the other hand, the NC phase factors in~\eqref{Helicity25} are momentum dependent and vanishes when then momenta invovled become parallel/antiparallel to each other. These NC phases, eventually, cancel all the collinear divergences from the Mandelstam variable fractions~\cite{Latas:2020nji}. Therefore, the collinear divergence distinguishes the NCQED scattering amplitudes from their QCD counterparts in this case.

\subsubsection{Collinear divergences of Compton scattering}

Like the gauge bosoon sector, the fermion-photon scattering amplitudes in NCQED also share some characters of the QCD quark-gluon scattering amplitudes~\cite{Godfrey:2001yy,Latas:2020nji}. Here it manifests as an additional term with $t^2$ denominator in additional to the Klein-Nishina formula of the massless amplitude square:
\begin{eqnarray}
{\sum\limits_{\rm spins}}\big\vert{\cal M}(\gamma e\to\gamma e)\big\vert^2
=8e^4\; \Bigg[-\frac{u}{s}-\frac{s}{u}
+4\frac{s^2+u^2}{t^2}\sin^2{\frac{k^{\gamma}_{in}\theta k^{\gamma}_{out}}{2}}\Bigg].
\label{NCampsquer}
\end{eqnarray}
Again, the collinear (forward scattering) singularity from $t^2$ denominator is softened from the QCD order by the NC factor in~\eqref{NCampsquer}. On the other hand, this time it becomes an  $t^{-1}$ divergenced instead of being completely cancelled as in the $\gamma\gamma\to\gamma\gamma$ process. This particular form of the collinear singularity in NCQED Compton scattering makes this theory different from both QED and QCD. Unlike the u-channel (back scattering) singularity in the Klein-Nishina formula, the forward t-channel scattering singularity in~\eqref{NCampsquer} is not regulated by the fermion mass. It is (still) an open question how this singularity, due to the t-channel photon propagator generated by the triple photon coupling, could be naturally regularized within NCQED~\cite{Latas:2020nji}.\footnote{We speculate that the answer to this question may be within/connected to the regularization of IR divergences in NCQED at loop levels, which is still unknown at this moment.}

\subsection{Moyal $\theta$-exact NCQED built by irreversible SW map -- second model}

Irreversible SW maps played a central role in the phenomenological NCYM model building. This approach allowed unlimited selections of non-Abelian gauge group/representation. It also enabled commutative matter fields with arbitrary U(1) charges to couple to the same commutative U(1) gauge field~\cite{Horvat:2011qn}. The $\theta$-exact quantum mechanical calculations in such models \cite{Martin:2020ddo,Trampetic:2021awu}, which should allow these theories to be compared with U(N) theories without SW map, have only been carried out in recent years because of their inherent tediousness. As an extension to the identity found for reversible SW map, a second NCQED model with two differently charged fermions coupled to the same commutative photon field was briefly investigated in~\cite{Trampetic:2021awu}, with the starting action being as follows:
\begin{equation}
\begin{split}
^2S\big[\Psi_1&(\psi_1, e a_\rho),\Psi_2(\psi_2, \kappa e a_\rho),
{A_1}_\mu(e a_\rho),{A_2}_\mu(\kappa e a_\rho)\big]
\\=\sum\limits_{i=1}^2&\int\bar\Psi_i \star(i\slashed{D}_{(i)}-m)\Psi_i
-\frac{1}{4G^2}\int \Big(F^2[{A_1}_\mu (e a_\rho)]
+\lambda F^2[{A_2}_\mu(\kappa e a_\rho)]\Big),
\\D_{(i)}^\mu\Psi_i
&=\partial^\mu\Psi_i-iA_i^\mu\star\Psi_i,
\;G^2=e^2(1+\lambda\kappa^2).
\label{SWAction}
\end{split}
\end{equation}
From this action, it is derived that the interaction vertex terms $^2V$ of our second model (\ref{SWAction}) can be expressed as nonidentical linear combinations of the first model \eqref{NCminAction} vertex terms $^1V$~\cite{Trampetic:2021awu}
\begin{gather}
\begin{split}
^2V_{\bar\psi_2 a^n\psi_2}&=\kappa^n \cdot{^2V_{\bar\psi_1 a^n\psi_1}}
=\kappa^n \cdot{^1V_{\bar\psi a^n\psi}},
\end{split}
\label{25}\\
\begin{split}
^2V_{(A_2\vert 2;i,j)}&={\lambda\kappa^{i+j}}\cdot\,^2V_{(A_1\vert 2;i,j)}
=\frac{\lambda\kappa^{i+j}}{1+\lambda\kappa^2}\;^1V_{(A\vert 2;i,j)},
\end{split}
\label{22}\\
\begin{split}
^2V_{(A_2\vert 3;i,j,k)}=&\lambda\kappa^{i+j+k}\cdot\,^2V_{(A_1\vert 3;i,j,k)}
=\frac{\lambda\kappa^{i+j+k}}{1+\lambda\kappa^2}\;^1V_{(A\vert 3;i,j,k)},
\end{split}
\label{23}\\
\begin{split}
& ^2V_{(A_2\vert 4;i,j,k,l)}
=\lambda\kappa^{i+j+k+l}\cdot\,^2V_{(A_1\vert 4;i,j,k,l)}
=\frac{\lambda\kappa^{i+j+k+l}}{1+\lambda\kappa^2}\;^1V_{(A\vert 4;i,j,k,l)},
\end{split}
\label{24}
\end{gather}
where $(A(A_i)\vert 2;i,j)$, $(A(A_i)\vert 3;i,j,k)$, $(A(A_i)\vert 4;i,j,k,l)$ denote fully ordered vertex terms derived from the quadratic, cubic, and quartic power of the NC field $A^\mu$ terms ($A_i^\mu$) in the action of the first model (\ref{NCminAction}) by $i,j$-, $i,j,k$-, and $i,j,k,l$-th order SW map, respectively~\cite{Trampetic:2021awu}.
Consequently, the identity between the first model scattering amplitudes with and without SW map no longer holds for the second model. Still, one can express the two-by-two scattering amplitudes of the second model as the first model terms plus SW map based correction terms $\Delta$:
\begin{gather}
\begin{split}
&^2\Gamma({\gamma\psi_1\to\gamma\psi_1})
{=^1}\Gamma({\gamma\psi\to\gamma\psi})
+\Delta({\gamma\psi_1\to\gamma\psi_1}),
\end{split}
\label{26}\\
\begin{split}
&^2\Gamma({\gamma\psi_2\to\gamma\psi_2})
{=\kappa^2\cdot^1}\Gamma({\gamma\psi\to\gamma\psi})
+\Delta({\gamma\psi_2\to\gamma\psi_2}),
\end{split}
\label{27}\\
\begin{split}
&^2\Gamma({\gamma\gamma\to\gamma\gamma})
{=^1}\Gamma({\gamma\gamma\to\gamma\gamma})
+\Delta({\gamma\gamma\to\gamma\gamma}),
\end{split}
\label{28}\\
\begin{split}
&\Delta({\gamma\psi_1\to\gamma\psi_1})=-\lambda\kappa\Delta({\gamma\psi_2\to\gamma\psi_2})
=\frac{\lambda\kappa^2(1-\kappa)}{1+\lambda\kappa^2}\Big[(\bar\Psi\slashed{A}\Psi;\bar 0,1,0)\Big]_{\rm on-shell}\,,
\end{split}
\label{29}\\
\begin{split}
&\Delta({\gamma\gamma\to\gamma\gamma})=\frac{\lambda^2\kappa^2(1-\kappa)^2}{(1+\lambda\kappa^2)^2}
\\&\times
\Big[(2;1,1)+(3;0,0,1)
+(3;0,1,0)+(3;1,0,0)\Big]_{\rm on-shell}.
\end{split}
\label{30}
\end{gather}
Furthermore, the SW map correction terms to the two-by-two scattering amplitudes of the second model are all vertex terms, therefore they do not add additional collinear divergences to the scattering amplitudes. Therefore, it is possible that the collinear divergences found in the first model are generic for all NCQED models.

\section{Particle phenomenology of Moyal NCQFT} 

As already mentioned, particle phenomenological applications of Moyal NC gauge theories first started with the models based on star products only, for example~\cite{Hewett:2000zp,Godfrey:2001yy}. More phenomenological NCQFT models were built shortly after using the SW map based approach. These models were usually expressed as an expansion to the first or second order of NC parameter $\theta$. The studies on their phenomenology (possible experimental signal/bounds on noncommutative background) started parallel to the pure theoretical developments of these models \cite{Schupp:2002up,Minkowski:2003jg}. The majority of the accelerator processes had been surveyed up to the second power in $\theta^{\mu \nu}$ \cite{Ohl:2004tn,Alboteanu:2006hh}.

Phenomenological applications of the $\theta$-exact SW maps~
\cite{Horvat:2010sr,Horvat:2011iv,Wang:2012ye,Horvat:2012vn,Selvaganapathy:2015nva,Horvat:2017gfm,Horvat:2017aqf,J.:2019bws,Horvat:2020ycy,Latas:2020nji,Bekli:2020unl,Trampetic:2023qfv} started a few years after it was used in loop calculation~\cite{Schupp:2008fs}. This approach turns out to be desirable from the phynomenological perspective too because the full $\theta$-exact NC factors can be shown to be finite for real momenta~\cite{Martin:2012aw} regardless the scales of $\theta$, which means that the resulting quantities are valid for much wider energy scales than those bounded from above by $\theta$~\cite{Horvat:2010sr,Horvat:2011iv,Horvat:2012vn}. Both, the accelerator physics and non-accelerator scenarios like astrophysics and cosmology have been studied. For identical scenarios, $\theta$-exact approach lead to the significant correction to the results based on leading orders of $\theta$-expansion if the process of interest takes place at energy scales similar to/higher than the NC scale.

U(1) NCYM can give rise to contributions to both known and unknown processes, since new couplings, of which the most distinctive being the various photon self-couplings and neutral particle like Z-boson and neutrinos(including sterile ones)-photon couplings, can now emerge in the noncommutative background even at the tree level.  Such couplings give rise to novel processes forbidden $(Z\to\gamma\gamma,\;\bar\nu_R\nu_R)$, or invisible $(Z\to\bar\nu_L\nu_L)$, in the SM \cite{Horvat:2010sr,Horvat:2011iv,Horvat:2012vn}. They also provide new channels in the known processes as Compton-, Møller-, Bhabha-like, fermion pair annihilations/production-like, $\gamma\gamma\to\gamma\gamma$ \cite{J.:2019bws,Horvat:2017gfm,Horvat:2017aqf,Horvat:2020ycy,Latas:2020nji}, and scatterings of ultra high energy (UHE) cosmogenic neutrinos --originating from the cosmic rays-- scattered on Earth nuclei  $(\nu +N\to\nu+anything)$ respectively \cite{Horvat:2010sr,Horvat:2011iv}.

Application of $\theta$-exact SW map method on the early universe right-handed neutrino physics considered in Ptolemy experiment \cite{Horvat:2017gfm,Horvat:2017aqf}, to disclose by it not only the decoupling temperature for the said neutrino component, but also the otherwise hidden coupling temperature.  Considering two relevant processes, the plasmon decay \cite{Horvat:2017gfm} and the neutrino-electron elastic scattering \cite{Horvat:2017aqf}, we study the interplay between the structures of the NC parameter $\theta^{\mu \nu}$ (the NC types) and the reheating temperature after inflation to obtain otherwise elusive upper bound on the scale $\Lambda_{\rm NC}$. If  Ptolemy enhanced capture rate is due to NC spacetime, we verify that a nontrivial maximum upper bound on $\Lambda_{\rm NC}$ (a way below the Planck scale) emerges for a space-like $\theta^{\mu \nu}$ and at sufficiently high reheating temperature.

Using the $\theta$-exact SW map and enveloping algebra formalisms, authors of \cite{J.:2019bws} constructed the Moyal $\theta$-exact NCSM, invariant under Cohen-Glashow Very Special Relativity Lorentz subgroup \cite{Cohen:2006ky}, and applied it to compute the high energy cross-section of the top quark pair production in $e^+ e^-$ collision for light-like noncomutativity.

Also very recently the first model was applied to specific 
$\gamma\gamma\to\gamma\gamma$ collisions, i.e. light-by-light scattering in vacuum, which is a nonlinear process forbidden at the classical tree level QED by the Landau-Young theorem, but predicted to occur in the SM via radiative quantum corrections. Light-by-light scattering is very interesting due to the recent ATLAS measurements \cite{Aad:2019ock} of $\rm PbPb\to Pb^*Pb^*\gamma\gamma$ collision, and it was used to constrain the first NCQED model in \cite{Horvat:2020ycy,Latas:2020nji}.

\section{Discussion and conclusion}

Motivated by various concepts from the string perspective and modifications of gravity (black holes), extenstive researches on spacetime noncommutativity have lasted over two decades. Recent progresses of one particullar branch, namely gauge theories on Moyal space as perturbative quantum field theories, are summarized in some details in this minireview. We mainly focus on the on-shell properties of noncommutative gauge theories defined via SW maps as quantum field theories. The equivalence of was first pointed out in Section 2. In Section 3 we went through some recent studies on tree level scattering amplitudes of NCQED after SW map, first the identity between tree level NCQED scattering amplitudes before and after reversible SW map, and then a revisit of the IR properties of the NCQED two-by-two scattering amplitudes. More general efforts on phenomenological effects of NCYM particle physics models are discussed in Section 4.

We consider the scattering amplitude identity as the manifestion of the formal equivalence mentioned in Section 2. The IR properties of two-by-two scattering amplitudes were not often mentioned in literature. On the other hand, the collinear divergence in NCQED Compton scattering amplitude may be a suggestion that instead of one loop, the IR structure of NCYM is nontrivial from tree level on. We would like to speculate that nontrivial (nonperturbative?) IR physics of NCQED may be necessary to solve the collinear divergence in Compton scattering and, hopefully, other IR divergences at loop levels too and wish for progresses alone this line in the near future. The forward scattering enhancement in NCQED Compton scattering \cite{Trampetic:2023qfv} may also be a good place to search for phenomenological effects if it could be properly regulated.


\acknowledgments{JT would like thank Dieter L\"ust for many discussions and to acknowledge support of Max-Planck-Institute for Physics, M\"unchen, Germany, for hospitality.}


\begin{thebibliography}{999}

\bibitem{ArkaniHamed:1998rs}
  Arkani-Hamed N., Dimopoulos S., Dvali G.R.,
  {\it The Hierarchy problem and new dimensions at a millimeter},
  Phys. Lett. B {\bf 429} (1998) 263, [arXiv:hep-ph/9803315].

\bibitem{AmelinoCamelia:1997gz}
  Amelino-Camelia G., Ellis J.R., Mavromatos N.E., Nanopoulos D.V., Sarkar S.,
 {\it Tests of quantum gravity from observations of gamma-ray bursts},
  Nature {\bf 393} (1998) 763. [arXiv:astro-ph/9712103].

\bibitem{Snyder:1946qz}
  H.~S.~Snyder, {\it Quantized space-time,} Phys.\ Rev.\  {\bf 71} (1947) 38, 
  doi:10.1103/PhysRev.71.38.
  
\bibitem{Snyder:1947nq}
  H.~S.~Snyder,
 {\it The Electromagnetic Field in Quantized Space-Time,} 
 Phys.\ Rev.\  {\bf 72} (1947) 68, doi:10.1103/PhysRev.72.68.

\bibitem{Seiberg:1999vs}
N. Seiberg and E. Witten,
{\it String theory and noncommutative geometry},
  JHEP {\bf 09} (1999) 032.
  
\bibitem{Kontsevich:1997vb}
  M.~Kontsevich,
 {\it Deformation quantization of Poisson manifolds. 1.} 
  Lett.\ Math.\ Phys.\  {\bf 66} (2003) 157,  doi:10.1023/B:MATH.0000027508.00421.bf
  [q-alg/9709040].

\bibitem{Madore:2000en}
  J.~Madore, S.~Schraml, P.~Schupp and J.~Wess,
  {\it Gauge theory on noncommutative spaces,}
  Eur.\ Phys.\ J.\ C {\bf 16} (2000) 161,  doi:10.1007/s100520050012
  [hep-th/0001203].
  
\bibitem{Jurco:2000fb}
  B.~Jurco and P.~Schupp,
 {\it Noncommutative Yang-Mills from equivalence of star products,}
  Eur.\ Phys.\ J.\ C {\bf 14} (2000) 367,  doi:10.1007/s100520000380
  [hep-th/0001032].
    
\bibitem{Jurco:2001rq}
  B.~Jurco, L.~Moller, S.~Schraml, P.~Schupp and J.~Wess,
  {\it Construction of nonAbelian gauge theories on noncommutative spaces,}
  Eur.\ Phys.\ J.\ C {\bf 21} (2001) 383,  doi:10.1007/s100520100731
  [hep-th/0104153].
  
\bibitem{Jurco:2001my}
  B.~Jurco, P.~Schupp and J.~Wess,
  {\it NonAbelian noncommutative gauge theory via noncommutative extra dimensions,}  
  Nucl.\ Phys.\ B {\bf 604} (2001) 148,  doi:10.1016/S0550-3213(01)00191-2, [hep-th/0102129].
        
\bibitem{Jurco:2001kp}
  B.~Jurco, P.~Schupp and J.~Wess,
 {\it Noncommutative line bundle and Morita equivalence,}
  Lett.\ Math.\ Phys.\  {\bf 61}, 171 (2002) [hep-th/0106110].
  
  \bibitem{Jackiw:2001jb}
  R.~Jackiw and S.~Y.~Pi,
  {\it Covariant coordinate transformations on noncommutative space},
  Phys.\ Rev.\ Lett.  {\bf 88} (2002) 111603,  [arXiv:hep-th/0111122].

\bibitem{Madore}
J. Madore, {\it An Introduction to Noncommutative Differential Geometry and its Physical Applications},
2nd Eddition (Cambridge University Press, 1999).

\bibitem{Gomis:2000zz}
  J.~Gomis and T.~Mehen,
{\it Space-time noncommutative field theories and unitarity,}
  Nucl.\ Phys.\  B {\bf 591}, 265 (2000),
  [arXiv:hep-th/0005129].

\bibitem{Aharony:2000gz}
  O.~Aharony, J.~Gomis, T.~Mehen,
On theories with lightlike noncommutativity,
JHEP{\bf 0009}, 023 (2000),  [hep-th/0006236].

\bibitem{Dirac:1928hu}
  P.~A.~M.~Dirac,
  {\it The quantum theory of the electron}, Proc.\ Roy.\ Soc.\ Lond.\ A {\bf 117} (1928) 610,  
  doi:10.1098/rspa.1928.0023.

\bibitem{Dirac:1931kp}
  P.~A.~M.~Dirac,
 {\it Quantised singularities in the electromagnetic field},
  Proc.\ Roy.\ Soc.\ Lond.\ A {\bf 133} (1931) no.821,  60,  
  doi:10.1098/rspa.1931.0130.

\bibitem{Born:1934gh}
  M.~Born and L.~Infeld,
{\it Foundations of the new field theory,}
 Proc.\ Roy.\ Soc.\ Lond.\ A {\bf 144} (1934) no.852,  425,  doi:10.1098/rspa.1934.0059.

\bibitem{Szabo:2009tn}
  Szabo R.J.,   {\it Quantum Gravity, Field Theory and Signatures of Noncommutative Spacetime},
  Gen. Rel. Grav.  {\bf 42} (2010)  1-29,   [arXiv:0906.2913].
  
\bibitem{Martin:1999aq}
C.P.~Martin, D.~Sanchez-Ruiz,
{\it The One-loop UV Divergent Structure of {\rm U(1)} Yang-Mills Theory on Noncommutative $R^4$},
Phys. Rev. Lett.  {\bf 83} (1999) 476--479, [hep-th/9903077].
  
\bibitem{Liu:2000mja}
  H.~Liu,
  {*-Trek II: *(n) operations, open Wilson lines and the Seiberg-Witten map,}
  Nucl.\ Phys.\ B {\bf 614} (2001) 305, [hep-th/0011125].
  
\bibitem{Okawa:2001mv}
  Y.~Okawa and H.~Ooguri,
{\it An Exact solution to Seiberg-Witten equation of noncommutative gauge theory,}
  Phys.\ Rev.\ D {\bf 64} (2001) 046009,  [hep-th/0104036].

\bibitem{Brace:2001rd}
  D.~Brace, B.~L.~Cerchiai and B.~Zumino,
{\it Nonabelian gauge theories on noncommutative spaces,}
  Int.\ J.\ Mod.\ Phys.\ A{\bf 17}, 205 (2002),  [hep-th/0107225].
  
\bibitem{Martin:2002nr}
C.~P. Martin,
{\it The gauge anomaly and the Seiberg-Witten map},
Nucl. Phys. {\bf B652} (2003) 72--92.

\bibitem{Brandt:2003fx}
F. Brandt, C.P. Martin and F. Ruiz Ruiz,
{\it Anomaly freedom in Seiberg-Witten noncommutative gauge theories},
JHEP {\bf 07} (2003) 068.
  
\bibitem{Barnich:2002pb}
  G.~Barnich, F.~Brandt and M.~Grigoriev,
  {\it Seiberg-Witten maps and noncommutative Yang-Mills theories for arbitrary gauge groups},
  JHEP {\bf 0208} (2002) 023.
 
 \bibitem{Martin:2012aw}
  C.~P.~Martin, {\it Computing the $\theta$-exact Seiberg-Witten map for arbitrary gauge groups},
  Phys.\ Rev.\ D {\bf 86} (2012) 065010, [arXiv:1206.2814 [hep-th]].
  
\bibitem{Trampetic:2015zma}
  J.~Trampetic and J.~You,
  {\it $\theta$-exact Seiberg-Witten maps at order $e^3$,}
  Phys.\ Rev.\ D {\bf 91} (2015) no.12,  125027,  doi:10.1103/PhysRevD.91.125027
  [arXiv:1501.00276 [hep-th]].
  
 \bibitem{Horvat:2011qn}
R.~Horvat, A.~Ilakovac, P.~Schupp, J.~Trampeti\'{c}, and J.~You,
{\it Yukawa couplings and seesaw neutrino masses in noncommutative gauge theory}, 
 {\em Phys. Lett.} {\bf B715}, 340 (2012),

  \bibitem{arXiv0711.2965B} 
M.~Bordemann, N.~Neumaier, S.~Waldmann, and S.~Weiss, 
{\it Deformation Quantization of Surjective Submersions and Principal Fibre Bundles},  
Crelle's J. reine angew. Math. 639 (2010), 1--38,
[Abstract] [PDF] [MR2608189] [Zbl05687061], arXiv:0711.2965.

\bibitem{arXiv0909.4259B} 
H.~Bursztyn, V.~Dolgushev, and S.~Waldmann, 
{\it Morita equivalence and characteristic classes of star products}, 
 Crelle's J. reine angew. Math. 662 (2012), 95-163, 
[Abstract] [PDF] [MR2876262] [Zbl1237.53080], arXiv:0909.4259.

\bibitem{Calmet:2001na}
X.~Calmet, B.~Jurco, P.~Schupp, J.~Wess, and M.~Wohlgenannt,
{\it The standard  model on non-commutative space-time},
 Eur. Phys. J. {\bf C23} (2002) 363--376.

\bibitem{Behr:2002wx}
W.~Behr, N.~Deshpande, G.~Duplan\v{c}i\'c, P.~Schupp, J.~Trampeti\'c, and J.~Wess,
  {\it The $Z \to \gamma \gamma, g g$ decays in the noncommutative standard model},  Eur. Phys. J. {\bf C29} (2003) 441--446.

\bibitem{Aschieri:2002mc}
P.~Aschieri, B.~Jurco, P.~Schupp, and J.~Wess,
 {\it Non-commutative GUTs, standard model and C, P, T},
 Nucl. Phys. {\bf B651} (2003) 45--70,

\bibitem{Martin:2013gma}
  C.~P.~Martin,
{\it The Minimal and the New Minimal Supersymmetric Grand Unified Theories on Noncommutative Space-time,}
 Class.\ Quant.\ Grav.\  {\bf 30} (2013) 155019.
  
   \bibitem{Buric:2006wm}
M.~Buric, V.~Radovanovic, and J.~Trampetic,
{\it The one-loop renormalization of the gauge sector in the noncommutative standard model},
JHEP {\bf  03} (2007) 030.

\bibitem{Latas:2007eu}
D.~Latas, V.~Radovanovic, and J.~Trampetic,
{\it Non-commutative SU(N) gauge theories and asymptotic freedom},
Phys. Rev. {\bf D76} (2007) 085006,
 
\bibitem{Buric:2007ix}
M.~Buric, D.~Latas, V.~Radovanovic and J.~Trampetic,
{\it The Absence of the 4 psi divergence in noncommutative chiral models,}
Phys. Rev. D \textbf{77}, 045031 (2008)
doi:10.1103/PhysRevD.77.045031
[arXiv:0711.0887 [hep-th]].

\bibitem{Martin:2009sg}
  C.~P.~Martin, C.~Tamarit,
  {\it Noncommutative GUT inspired theories and the UV finiteness of the fermionic four point functions,}
  Phys.\ Rev.\  {\bf D80}, 065023 (2009),  [arXiv:0907.2464].
  %
\bibitem{Martin:2009vg}
  C.~P.~Martin and C.~Tamarit,
  {\it Renormalisability of noncommutative GUT inspired field theories with anomaly safe groups,}
  JHEP {\bf 0912} (2009) 042,  [arXiv:0910.2677 [hep-th]].
    
\bibitem{Buric:2010wd}
  M.~Buric, D.~Latas, V.~Radovanovic and J.~Trampetic,
{\it Chiral fermions in noncommutative electrodynamics: renormalisability and dispersion,}  Phys.\ Rev.\  D {\bf 83} (2011) 045023,  arXiv:1009.4603 [hep-th].

\bibitem{Schupp:2002up}
P.~Schupp, J.~Trampetic, J.~Wess, and G.~Raffelt,
{\it {The photon neutrino  interaction in NC gauge field theory and astrophysical bounds}},
   Eur. Phys. J. {\bf C36} (2004) 405--410,

\bibitem{Minkowski:2003jg}
P.~Minkowski, P.~Schupp, and J.~Trampetic,
{\it Neutrino dipole moments and  charge radii in non- commutative space-time}, Eur. Phys. J. {\bf C37}  (2004) 123--128,

\bibitem{Hewett:2000zp}
  J.~L.~Hewett, F.~J.~Petriello and T.~G.~Rizzo,
 {\it Signals for noncommutative interactions at linear colliders},
  Phys.\ Rev.\ D {\bf 64} (2001) 075012,  [hep-ph/0010354].
  
\bibitem{Godfrey:2001yy}
  S.~Godfrey and M.~A.~Doncheski,
 {\it Signals for noncommutative QED in e gamma and gamma gamma collisions,}
  Phys.\ Rev.\ D {\bf 65} (2002) 015005, [hep-ph/0108268].
  
\bibitem{Garg:2011aa}
  S.~K.~Garg, T.~Shreecharan, P.~K.~Das, N.~G.~Deshpande and G.~Rajasekaran,
 {\it TeV Scale Implications of Non Commutative Space time in Laboratory Frame with Polarized Beams,}
  JHEP {\bf 1107} (2011) 024,  doi:10.1007/JHEP07(2011)024
  [arXiv:1105.5203].

\bibitem{Ohl:2004tn}
T.~Ohl and J.~Reuter,
{\it Testing the noncommutative standard model at a future photon collider},
 Phys. Rev. {\bf D70} (2004) 076007,

\bibitem{Alboteanu:2006hh}
A.~Alboteanu, T.~Ohl, and R.~Ruckl,
{\it Probing the noncommutative standard model at hadron colliders},
Phys. Rev. {\bf D74} (2006) 096004,

\bibitem{Buric:2007qx}
M.~Buric, D.~Latas, V.~Radovanovic, and J.~Trampetic,
{\it Nonzero $Z \to\gamma\gamma$ decays in the renormalizable gauge sector of the NCSM},
   Phys. Rev. {\bf D75} (2007) 097701,

   \bibitem{Mehen:2000vs}
T.~Mehen and M.~B. Wise,
{\it {Generalized *-products, Wilson lines and the solution of the Seiberg-Witten equations}}, 
JHEP {\bf 12} (2000) 008,  

\bibitem{Grosse:2005iz}
  H.~Grosse and M.~Wohlgenannt,
  {\it On $\kappa$-deformation and UV/IR mixing,}
  Nucl.\ Phys.\ B {\bf 748} (2006) 473,  doi:10.1016/j.nuclphysb.2006.05.004
  [hep-th/0507030].

\bibitem{Meljanac:2011cs}
  S.~Meljanac, A.~Samsarov, J.~Trampetic and M.~Wohlgenannt,
  {\it Scalar field propagation in the $\phi^4$ kappa-Minkowski model,}
  JHEP {\bf 12} (2011) 010.
  
\bibitem{Meljanac:2017grw}
  S.~Meljanac, S.~Mignemi, J.~Trampetic and J.~You,
 {\it Nonassociative Snyder $\phi^4$ Quantum Field Theory,}
  Phys.\ Rev.\ D {\bf 96} (2017) no.4,  045021,  doi:10.1103/PhysRevD.96.045021 [arXiv:1703.10851 [hep-th]].
  
\bibitem{Meljanac:2017jyk}
  S.~Meljanac, S.~Mignemi, J.~Trampetic and J.~You,
{\it UV-IR mixing in nonassociative Snyder $\phi^4$ theory,}
  Phys.\ Rev.\ D {\bf 97} (2018) no.5,  055041,  doi:10.1103/PhysRevD.97.055041
  [arXiv:1711.09639 [hep-th]].

\bibitem{Filk:1996dm}
T.~Filk,
{\it Divergencies in a field theory on quantum space,}
Phys. Lett. B \textbf{376} (1996), 53-58
doi:10.1016/0370-2693(96)00024-X.

\bibitem{Minwalla:1999px}
  S.~Minwalla, M.~Van Raamsdonk and N.~Seiberg,
 {\it Noncommutative perturbative dynamics,}
  JHEP {\bf 0002}, 020 (2000),  [arXiv:hep-th/9912072].
  
  \bibitem{Hayakawa:1999yt}
M.~Hayakawa,
{\it Perturbative analysis on infrared aspects of noncommutative QED on  R**4,} 
Phys.\ Lett.\  B {\bf 478}, 394 (2000),  [arXiv:hep-th/9912094].
  
\bibitem{VanRaamsdonk:2000rr}
  M.~Van Raamsdonk and N.~Seiberg,
{\it Comments on noncommutative perturbative dynamics,} JHEP {\bf 0003} (2000) 035, doi:10.1088/1126-6708/2000/03/035, [hep-th/0002186].

\bibitem{Matusis:2000jf}
A.~Matusis, L.~Susskind, and N.~Toumbas,
{\it The IR/UV connection in the  non-commutative gauge theories},
 {\em JHEP} {\bf 12} (2000) 002,

\bibitem{VanRaamsdonk:2001jd}
M.~Van Raamsdonk,
{\it The Meaning of infrared singularities in noncommutative gauge theories,'}
JHEP \textbf{11} (2001), 006, doi:10.1088/1126-6708/2001/11/006 [arXiv:hep-th/0110093 [hep-th]].

\bibitem{Horvat:2011bs}
  R.~Horvat, A.~Ilakovac, J.~Trampetic and J.~You,
{\it On UV/IR mixing in noncommutative gauge field theories,}
JHEP {\bf 12} (2011) 081, arXiv:1109.2485 [hep-th].
  
\bibitem{Ferrari:2003vs}
A.~F.~Ferrari, H.~O.~Girotti, M.~Gomes, A.~Y.~Petrov, A.~A.~Ribeiro, V.~O.~Rivelles and A.~J.~da Silva,
{\it Superfield covariant analysis of the divergence structure of noncommutative supersymmetric QED(4),}
Phys. Rev. D \textbf{69} (2004), 025008
doi:10.1103/PhysRevD.69.025008
[arXiv:hep-th/0309154 [hep-th]].

\bibitem{Ferrari:2004ex}
A.~F.~Ferrari, H.~O.~Girotti, M.~Gomes, A.~Y.~Petrov, A.~A.~Ribeiro, V.~O.~Rivelles and A.~J.~da Silva,
{\it Towards a consistent noncommutative supersymmetric Yang-Mills theory: Superfield covariant analysis,}
Phys. Rev. D \textbf{70} (2004), 085012
doi:10.1103/PhysRevD.70.085012
[arXiv:hep-th/0407040 [hep-th]].

\bibitem{Horvat:2010km}
  R.~Horvat, J.~Trampetic,
 {\it Constraining noncommutative field theories with holography,}
  JHEP {\bf 1101 } (2011)  112,  [arXiv:1009.2933].
  
\bibitem{Lust:2017wrl}
  D.~Lust and E.~Palti,
 {\it Scalar Fields, Hierarchical UV/IR Mixing and The Weak Gravity Conjecture,}
  JHEP {\bf 1802} (2018) 040, doi:10.1007/JHEP02(2018)040  [arXiv:1709.01790].
  
\bibitem{Martin:2017nhg}
  C.~P.~Martin, J.~Trampetic and J.~You,
 {\it Quantum noncommutative ABJM theory: first steps,}
  JHEP {\bf 1804} (2018) 070,  doi:10.1007/JHEP04(2018)070
  [arXiv:1711.09664].
     
\bibitem{Zeiner:2007}
J.~Zeiner, {\it Noncommutative quantumelectrodynamics from Seiberg-Witten Maps
to all orders in Theta(mu nu)} (Wurzburg U.). Jul 2007. 139 pp.
\newblock PhD thesis.

  \bibitem{Schupp:2008fs}
P.~Schupp and J.~You,
{\it {UV/IR mixing in noncommutative QED defined by  Seiberg-Witten map}},
  JHEP {\bf 08} (2008) 107,

\bibitem{Martin:2016zon}
  C.~P.~Martin, J.~Trampetic and J.~You,
  {\it Super Yang-Mills and $\theta$-exact Seiberg-Witten map: absence of quadratic noncommutative IR divergences,}
  JHEP {\bf 1605} (2016) 169,  doi:10.1007/JHEP05(2016)169
  [arXiv:1602.01333 [hep-th]].
  
\bibitem{Horvat:2011qg}
R.~Horvat, A.~Ilakovac, P.~Schupp, J.~Trampetic and J.~You,
{\it Neutrino propagation in noncommutative spacetimes,}
JHEP \textbf{04} (2012), 108
doi:10.1007/JHEP04(2012)108
[arXiv:1111.4951 [hep-th]].
    
\bibitem{Horvat:2013rga}
R.~Horvat, A.~Ilakovac, J.~Trampetic and J.~You,
{\it`Self-energies on deformed spacetimes,}
JHEP \textbf{11} (2013), 071
doi:10.1007/JHEP11(2013)071
[arXiv:1306.1239 [hep-th]].

\bibitem{Horvat:2015aca}
R.~Horvat, J.~Trampeti\'c and J.~You,
{\it Photon self-interaction on deformed spacetime,}
Phys. Rev. D \textbf{92} (2015) no.12, 125006
doi:10.1103/PhysRevD.92.125006
[arXiv:1510.08691 [hep-th]].

\bibitem{Martin:2016saw}
  C.~P.~Martin, J.~Trampetic and J.~You,
  {\it Quantum duality under the $\theta$-exact Seiberg-Witten map,}
  JHEP {\bf 1609} (2016) 052,  doi:10.1007/JHEP09(2016)052
  [arXiv:1607.01541].
  
\bibitem{Martin:2016hji}
  C.~P.~Martin, J.~Trampetic and J.~You,
 {\it Equivalence of quantum field theories related by the $\theta$-exact Seiberg-Witten map,}
  Phys.\ Rev.\ D {\bf 94} (2016) no.4,  041703,  doi:10.1103/PhysRevD.94.041703
  [arXiv:1606.03312 [hep-th]].
  
\bibitem{Latas:2020nji}
D.~Latas, J.~Trampeti\'c and J.~You,
{\it Seiberg-Witten map Invariant Scatterings,}
Phys. Rev. D \textbf{104} (2021) no.1, 015021, doi:10.1103/PhysRevD.104.015021,
[arXiv:2012.07891].

\bibitem{Horvat:2020ycy}
R.~Horvat, D.~Latas, J.~Trampeti\'c and J.~You,
{\it Light-by-Light Scattering and Spacetime Noncommutativity,}
Phys. Rev. D \textbf{101} (2020) no.9, 095035
doi:10.1103/PhysRevD.101.095035, [arXiv:2002.01829 [hep-ph]].

\bibitem{Trampetic:2021awu}
J.~Trampeti\'c and J.~You,
{\it Seiberg-Witten maps and scattering amplitudes of NCQED,}
Phys. Rev. D \textbf{105} (2022) no.7, 075016, 
doi:10.1103/PhysRevD.105.075016, [arXiv:2111.04154 [hep-th]].

\bibitem{Raju:2009yx}
  S.~Raju, {\it The Noncommutative S-Matrix,}
  JHEP {\bf 0906} (2009) 005,  doi:10.1088/1126-6708/2009/06/005.

\bibitem{Huang:2010fc}
  J.~H.~Huang, R.~Huang and Y.~Jia,
  {\it Tree amplitudes of noncommutative U(N) Yang-Mills Theory,}
  J.\ Phys.\ A {\bf 44} (2011) 425401,  doi:10.1088/1751-8113/44/42/425401.
  
\bibitem{Peskin:1995ev}
  M.~E.~Peskin and D.~V.~Schroeder,
 {\it An Introduction to quantum field theory,}
 PERSEUS BOOKS, Cambridge, Massachusetts, USA.
  
\bibitem{Martin:2020ddo}
C.~P.~Martin, J.~Trampeti\'c and J.~You,
{\it UV/IR mixing in noncommutative SU(N) Yang\textendash{}Mills theory,}
Eur. Phys. J. C \textbf{81} (2021) no.10, 878, doi:10.1140/epjc/s10052-021-09686-5, 
[arXiv:2012.09119 [hep-th]].
           
\bibitem{Horvat:2010sr}
  R.~Horvat, D.~Kekez, J.~Trampetic,
 {\it Spacetime noncommutativity and ultra-high energy cosmic ray experiments,}
  Phys.\ Rev.\  {\bf D83 } (2011)  065013,  [arXiv:1005.3209 [hep-ph]].

\bibitem{Horvat:2011iv}
  R.~Horvat, D.~Kekez, P.~Schupp, J.~Trampetic, J.~You,
 {\it Photon-neutrino interaction in theta-exact covariant noncommutative field theory,} 
 Phys. Rev. {\bf D84} (2011) 045004, 

\bibitem{Horvat:2012vn}
  R.~Horvat, A.~Ilakovac, D.~Kekez, J.~Trampetic and J.~You,
{\it Forbidden and Invisible Z Boson Decays in Covariant theta-exact Noncommutative Standard Model},  J.\ Phys.\ G: Nucl. Part. Phys. {\bf 41} (2014) 055007, arXiv:1204.6201 [hep-ph]. 
  
\bibitem{Horvat:2017gfm}
  R.~Horvat, J.~Trampetic and J.~You,
  {\it Spacetime Deformation Effect on the Early Universe and the PTOLEMY Experiment,}  Phys.\ Lett.\ B {\bf 772} (2017) 130,  doi:10.1016/j.physletb.2017.06.028 [arXiv:1703.04800 [hep-ph]].

\bibitem{Horvat:2017aqf}
  R.~Horvat, J.~Trampetic and J.~You,
  {\it Inferring type and scale of noncommutativity from the PTOLEMY experiment,}  Eur.\ Phys.\ J.\ C {\bf 78} (2018) no.7,  572,  doi:10.1140/epjc/s10052-018-6052-1 [arXiv:1711.09643 [hep-ph]].
  
\bibitem{J.:2019bws}
  J.~Selvaganapathy, P.~Konar and P.~K.~Das,
{\it Inferring the covariant $\Theta$-exact noncommutative coupling in the top quark pair production at linear colliders},  
  JHEP {\bf 1906} (2019) 108,  doi:10.1007/JHEP06(2019)108 [arXiv:1903.03478 [hep-ph]].
  
\bibitem{Wang:2012ye}
W.~Wang, J.~H.~Huang and Z.~M.~Sheng,
{\it TeV Scale Phenomenology of $e^+e^- \to\mu^+ \mu^-$ Scattering in the Noncommutative Standard Model with Hybrid Gauge Transformation,}
Phys. Rev. D \textbf{86} (2012), 025003
doi:10.1103/PhysRevD.86.025003
[arXiv:1205.0666 [hep-ph]].

\bibitem{Selvaganapathy:2015nva}
J.~Selvaganapathy, P.~K.~Das and P.~Konar,
{\it Search for associated production of Higgs with Z boson in the noncommutative Standard Model at linear colliders,}
Int. J. Mod. Phys. A \textbf{30} (2015) no.26, 1550159
doi:10.1142/S0217751X15501596
[arXiv:1509.06478 [hep-ph]].

\bibitem{Bekli:2020unl}
M.~R.~Bekli, I.~Chadou and N.~Mebarki,
{\it Bounds on the scale of noncommutativity from mono photon production in ATLAS Runs -1 and -2 experiments at LHC energies,} 
Int. J. Geom. Meth. Mod. Phys. \textbf{18} (2021) no.08, 2150126
doi:10.1142/S0219887821501267
[arXiv:2012.04331 [hep-ph]].

\bibitem{Trampetic:2023qfv}
J.~Trampeti\'c and J.~You,
{\it High energy Compton scattering in the NCQED,}
[arXiv:2301.06343 [hep-ph]].

\bibitem{Cohen:2006ky}
  A.~G.~Cohen and S.~L.~Glashow,
 {\it Very special relativity,}
  Phys.\ Rev.\ Lett.\  {\bf 97} (2006) 021601
  doi:10.1103/PhysRevLett.97.021601, [hep-ph/0601236].

\bibitem{Aad:2019ock}
  G.~Aad {\it et al.} [ATLAS Collaboration],
 {\it Observation of light-by-light scattering in ultraperipheral Pb+Pb collisions with the ATLAS detector,}  Phys.\ Rev.\ Lett.\  {\bf 123} (2019) no.5,  052001, doi:10.1103/PhysRevLett.123.052001, [arXiv:1904.03536 [hep-ex]].
      

\end{thebibliography}
\end{document}